%% file: asas3.tex
\documentclass[times,11pt]{article}
\input{include.tex}

\newcommand{\FigurePs}[7]{\begin{figure}[htb]\vspace{#1}
\includegraphics{#4}
\FigCap{#2}\label{#3}
\end{figure}}
\newcommand{\FigurePsSep}[7]{\begin{figure}[p]\vspace{#1}
\includegraphics{#4}
\FigCap{#2}\label{#3}
\end{figure}}
 
\begin{document}
\def\thefootnote{\fnsymbol{footnote}}
\begin{Titlepage}
\Title{The All Sky Automated Survey. A Catalog of almost 3900 variable stars.
\footnote{Based on observations obtained
at the Las Campanas Observatory of the Carnegie Institution of Washington.}}
\Author{G.~~P~o~j~m~a~\'n~s~k~i}{Warsaw University Observatory
Al~Ujazdowskie~4, 00-478~Warsaw, Poland\\
e-mail: gp@sirius.astrouw.edu.pl}
 
\end{Titlepage}

\vspace*{-12pt}
\Abstract{Results of the first two years of observations using the All Sky
Automated Survey prototype camera are presented. More than 140 000 stars
in 50 Selected Fields covering 300 sq. degrees were monitored each clear night
in the $I$-band resulting in the ASAS Photometric $I$-band Catalog  containing 
over $50 \times 10^6$ individual measurements. Nightly monitoring
of over 100 standard stars confirms that most of our data remains within
$\sigma_I=0.03$ of the standard $I$ system.
Search for the stars varying on the time scales 
longer than a day revealed almost 3900 variables (mostly irregular,
pulsating and binaries) brighter than 13 mag. 
Only 167 of them are known variables included in GCVS, 56 were 
observed by Hipparcos satellite (46 were marked as variable).
Among the stars brighter than $I \sim 7.5$ (which are saturated
on our frames) we have found about 50 variables (12 are in GCVS, 6 other in 
Hipparcos (Perryman \etal 1997) catalog). 
Because of the large volume of the data we present here only
selected tables and light-curves, but the complete ASAS Catalog of Variable 
Stars (currently divided into Periodic and Miscellaneous sections)
and all photometric data are available on the Internet
{\em http://www.astrouw.edu.pl/$\sim$gp/asas/asas.html} or 
{\em http://archive.princeton.edu/$\sim$asas/}
}{Catalogs -- Stars:variables -- Surveys} 

\vspace*{-6pt} 
\Section{Introduction}
The All Sky Automated Survey (Pojmañski 1997, hereafter Paper I) 
is a low-cost observing project dedicated to detection and investigation of 
photometric variability of stars and other objects all over the sky 
(Paczy{\'n}ski 1997). 
We have started in 1997 with the small, automated prototype instrument
equipped with 768 $\times$ 512 MEADE Pictor~416 CCD camera, 135~mm
f/1.8 telephoto lens and $I$-band (Schott RG-9, 3mm) filter.
The limiting magnitude of the system is 13.  The instrument
was placed at the Las Campanas Observatory which is operated by
the Carnegie Institution of Washington.  LCO kindly allocated space
in the slide-roof house of the 10" astroghraph, and 
OGLE-2 team (Udalski, Kubiak and Szymañski 1997) generously
helps us making the system running.

Detailed description of the prototype instrument, data acquisition and 
reduction process and the ASAS Catalog were described in Paper I
and the first results of the search for short time scale periodic variables were
presented by Pojmañski (1998, hereafter Paper II).

This paper presents results of the search for long term  variables in
the ASAS Selected Fields using data obtained during the first two years of the
prototype instrument operation, monitoring 300 square degrees of the sky.
Although many short-term periodic 
variables have been discovered in the course of current work, 
detailed search for the variables with periods shorter than 1 day is still to
be done.

\FigurePs{10cm}
{Standard deviation $\sigma_I$ of the stellar magnitudes {\em vs} $I$-band 
magnitudes for 140000 stars observed so far. Filled dots and open circles 
represent miscellaneous and periodic variables respectively.
}
{fig1}{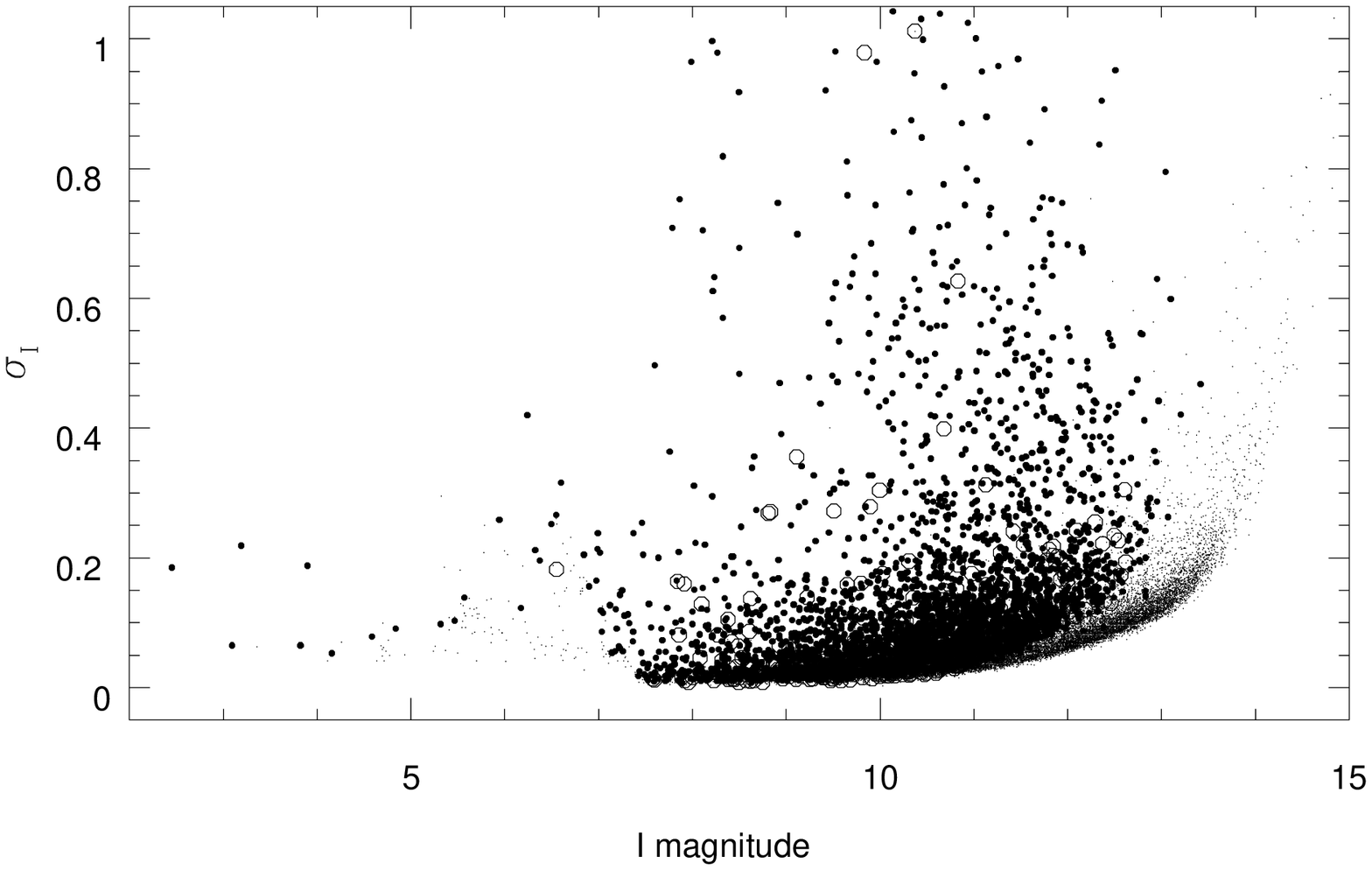}{70}{-30}{0}

\Section{Observations, Data Reduction, Variability Search }

\FigurePs{6.5cm}
{Difference between a) Hipparcos (Perryman \etal 1997)  "$I$" entries,
b) standard Landolt (1992) $I$ magnitudes and
 ASAS $I-band$ measurements for common stars.
}
{fig2}{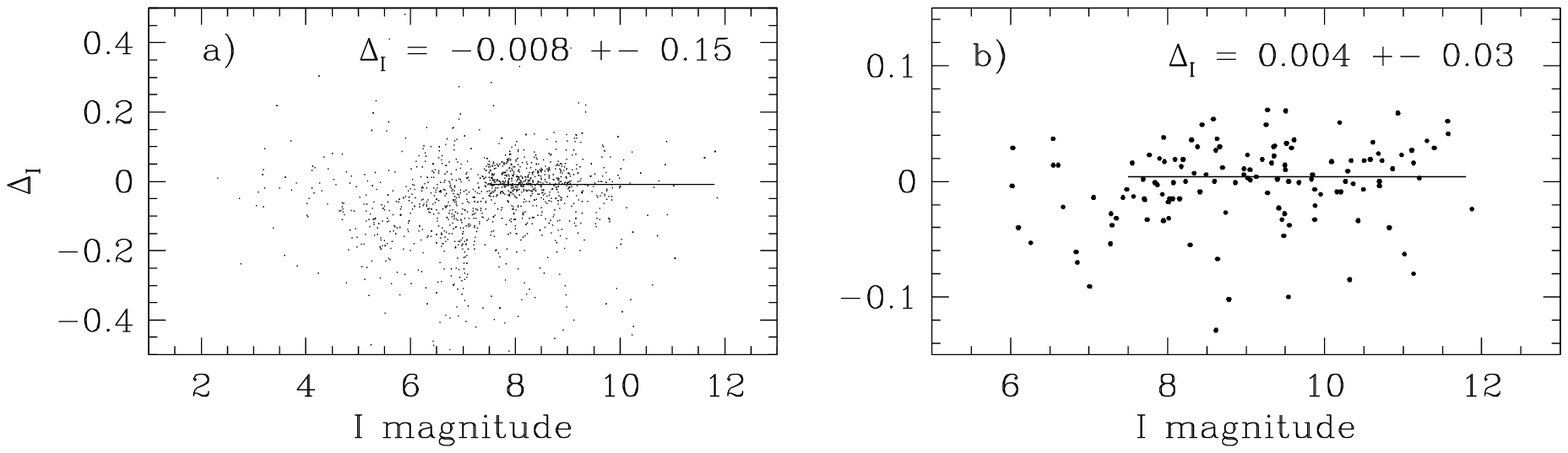}{67}{-30}{-30}

\FigurePs{6cm}
{Landolt (1992) stars found to be variable in the ASAS data.
}
{fig3}{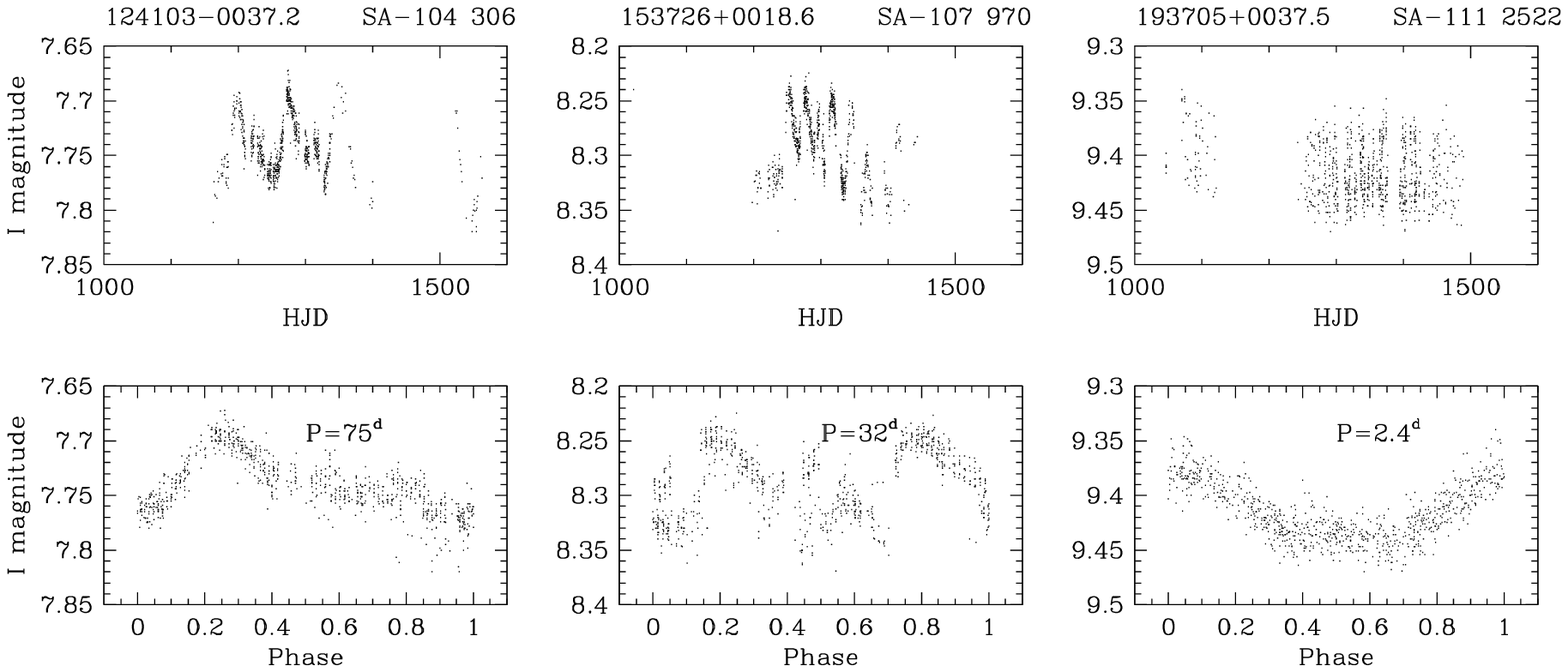}{60}{-10}{-30}

Fifty  2 $\times$ 3 deg fields (24 old, described in Paper II, and 26
new, mostly Landolt (1992)) were observed. Old fields were observed only
once per night, while the new ones - initially more often than that.
Coordinates of the Selected Field centers
as well as number of observed stars $N_{star}$,
number of detected miscellaneous
$N_{misc}$ and periodic $N_{per}$ variables and corresponding 
detection rates $f_{misc} = N_{misc}/N_{star}$ and $f_{per} = N_{per}/N_{star}$
respectively are given in Table \ref{table1}. Table \ref{table2} 
lists numbers of newly detected  and previously known variables 
and total number of observed stars in
1 magnitude bins.

\input table1.tex
\input table2.tex
\input table3.tex
\input table4.tex

Data reduction process was similar to this presented in Paper II. There
were however two differences. The flatfielding process was slightly
improved, so the magnitudes in the overlapping regions of different frames
are much closer to each other (usually stay within 0.05 mag). Photometry
was related to the  $I$-band magnitudes compiled by Hipparcos 
(Perryman \etal 1997) project.
Only data with $I$ magnitudes determined using methods
A-H (transformed direct measurements) were used, since most of the others
displayed large scatter when compared to our data (Fig. \ref{fig2}a). 
The final check of
our photometry was done on 104 unsaturated standards measured in
24 Landolt (1992) equatorial fields. Only for 10 stars 
the difference between
our $I$-band and standard $I$ magnitudes was larger than 0.1 mag. 
For the remaining 94 stars (Fig. \ref{fig2}b) the average calibration error is 
$\Delta I = 0.004 \pm 0.03$. 3 Landolt (1992) standards were found to be
variable. Their light curves are presented in Fig. \ref{fig3}.

We believe therefore, that for most of the stars ASAS measurements are 
close to the standard system and, since measured in the uniform way, provide
more accurate $I$-band data than Hipparcos (Perryman \etal 1997) compilation. 
One should be aware however that we
have performed aperture photometry only, so for stars having close companions
($ < 1 $ arcmin = 4 pixels) results are systematically biased.

The final data base - The ASAS Photometric $I$-band Catalog (containing over
140,000 stars and over $50\times 10^6$ individual measurements) is accessible
over the Internet.

In our analysis we have also included saturated stars. This was done with
dedicated routines measuring bleeding regions. For most stars we were able
to keep an {\it rms} error below the 0.1 mag level, but in some cases
(mostly for stars with $5.5 < I < 7.5$ for which presence of saturation depends
on the observing conditions) the error
was much larger. Justification for such effort
was the final score: over 40 new bright variables, but since the mean 
magnitudes of the saturated stars were brought closer to the standard system
by non-linear transformations - they should be used with extreme caution.

\FigurePsSep{20cm}
{The first 40 periodic variables from the ASAS Catalog of Variable Stars.
}
{fig4}{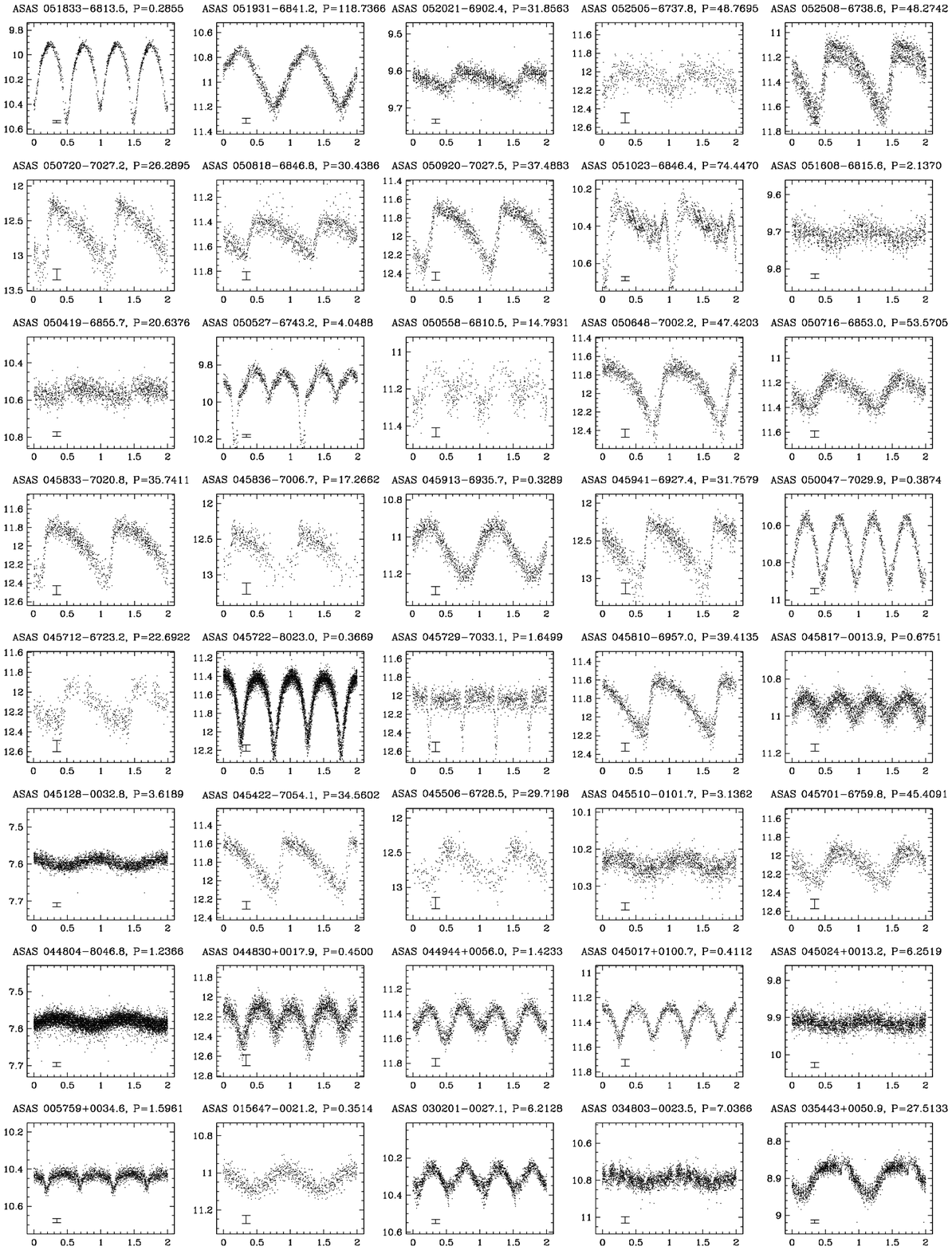}{80}{-60}{-60}

\FigurePsSep{20cm}
{Representation of the mscellaneous variables in the ASAS Catalog Variable Starss.
}
{fig5}{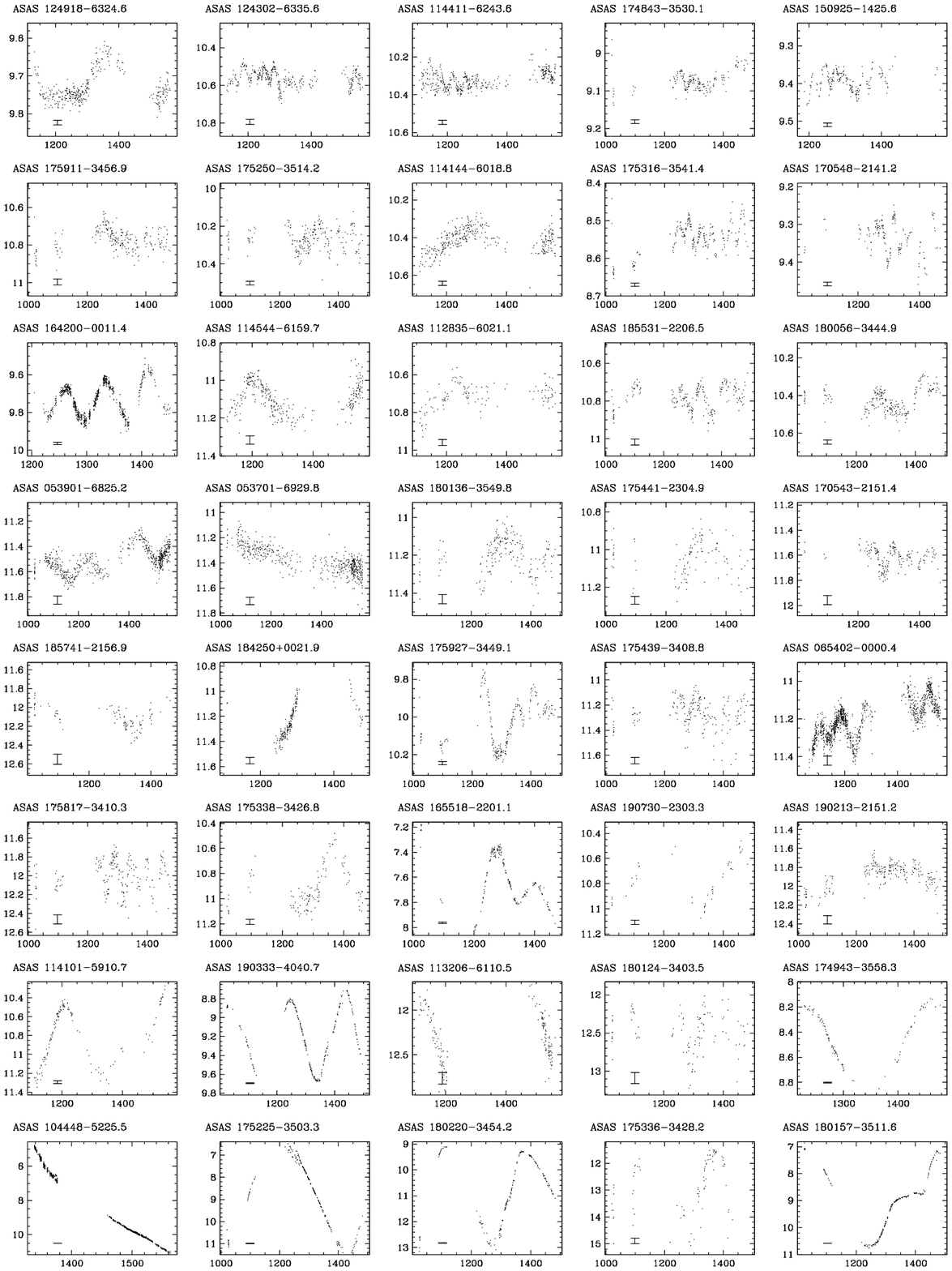}{80}{-60}{-60}

\FigurePsSep{20cm}
{Selected stars from the ASAS Catalog of Variable Stars.
}
{fig6}{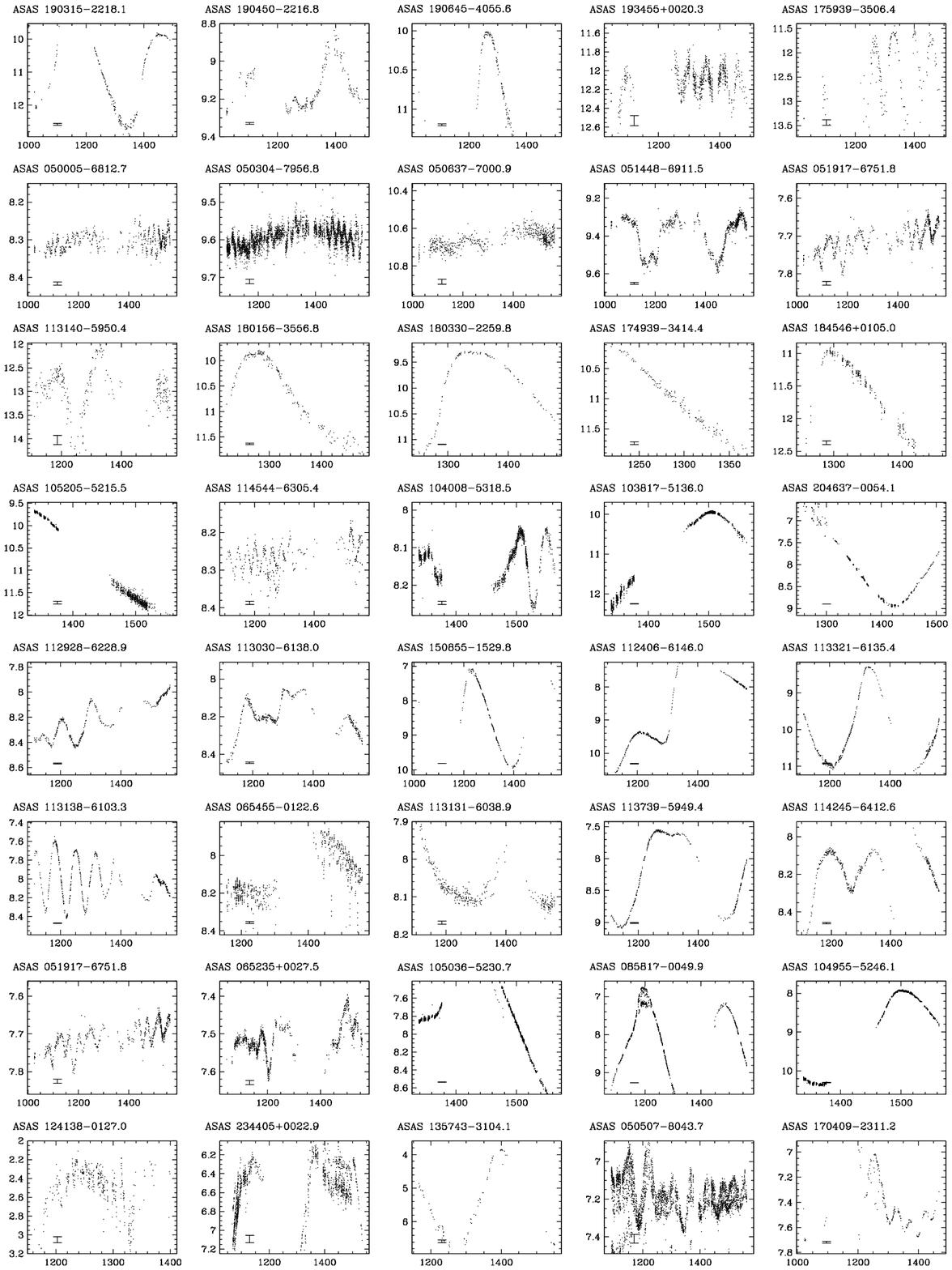}{80}{-60}{-60}

The detection algorithm which we have used was optimized for the
long-term variables.
First, the standard deviations $\sigma_I$ 
of stellar magnitudes were plotted against
the $I$-band magnitudes (Fig. \ref{fig1}), and the lower envelope of the 
plotted points was determined. Stars with magnitude dispersion larger then
$2\sigma$ were selected and their light curves extracted from the catalog.
Running median was than subtracted from the light curve and
new magnitude dispersion calculated. If it became significantly reduced
(3 times or more) light curve was inspected by eye. Otherwise
light curve  was tested for periodicity using the analysis of variance
(AoV) method (Schwarzenberg-Czerny 1989). The minimum frequency used in this 
calculation was 1 day - variables with shorter periods were detected only
if they contained enough power also at lower frequencies. 

Initially over 6000 light curves passed the long-term variability test and
another 1000 - period detection test. This set contained a lot of 
"garbage" mainly due to a) full-moon interference in ecliptic and equatorial 
fields, b) annual error for circumpolar fields, c) small number of measurements,
d) one-day observing cycle aliasing,
f) photometric errors for close companions,
e) other system inaccuracies.

Trying to do our best we have removed most of the spurious variables by hand.
It must be understood however, that some residual
garbage (mainly low-amplitude) might have survived.
Also data for saturated stars ($I<7.5$) are far from what we would
like to present - it should be used only as an indication of the 
variability.

\Section{Results}

The final ASAS Catalog of Variable Stars contains almost 3900 stars.
Currently it is divided in 2 sections:  a) Periodic Variables 
(binaries and pulsating stars with well defined periods) 
containing about 385 objects and
b) Miscellaneous Variables 
(quasi-periodic, irregular, slow pulatsors, multi-periodic and other) 
with over 3500 objects which will be successively classified and 
appointed to the different sections in the Catalog.

All detected variables were searched for in the GCVS (Kholopov 1985)
 and Hipparcos (Perryman \etal 1997) databases.
Only 47 out of 385 periodic and 120 out of 3500  miscellaneous variables are
included in GCVS. Hipparcos includes 19 (11 with periods listed) periodic 
objects and 37 miscellaneous (9 classified as non-variable).

Because of the volume of the data, only small subset of our results 
could be included in this paper.
The most prominent miscellaneous variables 
($I<12^{\rm m}$ and $\Delta I >1.5$) are
listed in Table \ref{table3} and a subset of
periodic variables ($I<12^{\rm m}$ and $\Delta I >0.3$)
in Table \ref{table4}.

Some examples of the available light curves are also presented in 
Figs. \ref{fig4}-\ref{fig6}.
Fig. \ref{fig4} shows the first 40 light curves taken from the catalog of
periodic variables.
Fig. \ref{fig5} contains light curves uniformly selected from the catalog 
of the miscellaneous variables and sorted according to the amplitude of variation.
Fig. \ref{fig6} is a subjective selected appetizer presenting some 
interesting light curves.

\Section {The ASAS Catalogs - Internet Access}

Complete versions of the ASAS Catalog of Variable Stars and
the ASAS Photometric $I$-band Catalog are available over the Internet.

The ASAS Home Page is located at: 
{\em http://www.astrouw.edu.pl/$\sim$gp/asas/asas.html} \\
and its US mirror at:
{\em http://archive.princeton.edu/$\sim$asas/}

The following services are available (please refer to the ASAS Home Page
to obtain correct URL addresses):
\begin{itemize}
\item{ASAS Catalog of Variable Stars containing lists of Miscellaneous and  Periodic Variables with}
  \begin{itemize}
    \item{ID (ASAS identification constructed using $\alpha_{2000}$ and $\delta_{2000}$ coordinates)}
    \item{$I$-band mean brightness}
    \item{Amplitude of variation}
    \item{Number of available measurements}
    \item{Period or time-scale of variation}
    \item{Cross-identification}
  \end{itemize}
  \item{Gallery of Periodic and Miscellaneous Variables' light curves (over 120 pages, 32 light curves each)}
  \item{ASAS Photometric $I$-band Catalog Interface providing access to full
photometric data (HJD, $I$) for}
\begin{itemize}
  \item{individual objects selected by ID or coordinates}
  \item{lists of objects}
  \item{objects in the specified box}
 \end{itemize}
  \item{ASAS Sky Atlas providing interactive maps of objects in the Catalog}
  \item{Download Page with}
  \begin{itemize}
    \item{ASAS Catalog of Variable Stars - lists of miscellaneous and periodic stars (53 kB and 8 kB) and compressed light curves (1.7 MB and 12MB)}
    \item{ASAS Photometric $I$-band Catalog - (ID, mean-$I$, $\sigma_I$) for all available objects (about 140,000 entries, 1.4 MB)} 
  \end{itemize}
\end{itemize}

\Section{Summary}

After two years of operation the ASAS Catalog of Variable Stars
contains now almost 3900 objects found among 140000 stars on
300 sq. degrees.
Although most of the stars are bright - 
only 5\% were previously known variables.
Most of new detections are bright, red irregular variables, for which some
classification schemes are required.
Among the periodic stars there are many examples of bright (7-9 magnitude)
Algol and WUMa binaries - easy targets for small size spectroscopic
instruments.

It is striking that even among very bright stars, saturated on our small CCD 
camera, a large fraction of variables has not been discovered yet. Note that new discoveries include lots of stars with large amplitudes as listed in 
Tables 3 and 4. Also note, that the 3900 variables were found by this 
pilot project in just 0.7\% of the whole sky.

The prototype instrument, consisting of the simple off-the-shelf amateur camera,
telephoto lens and custom made equatorial mount proved to be a very
robust and efficient tool. It is perfectly well suited  for detecting
and monitoring variability on the sky at a very low cost, thanks to
a high level of automation and simplicity of equipment. It required
only two author's visits to Las Campanas 
(4 weeks in April 1997 and 4 weeks in July 1998) to set up the system 
and let it run for over 2 years. 
The OGLE team provided regular (1 per month) tape
exchange service and occasional rebooting of the system (necessary due to 
rare and hard to pin down flaws in software).

In the next months
the prototype instrument will be replaced by two larger (2K x 2K) 
devices, capable to monitor 20000 sq. degrees per night. We expect
to increase our data steam by a factor of 30, so any achievements
in the field of automated variability  detection and 
light curve classification will be highly appreciated. We encourage 
interested people to use our data for testing purposes.

\vspace*{-6pt} 
\Acknow{
This project was made possible by a generous gift from Mr. William Golden
to Dr. Bohdan Paczy\'nski, and funds from Princeton University.  It is a
great pleasure to thank Dr. B. Paczy\'nski for his initiative, interest,
valuable discussions, and the funding of this project.

I am indebted to the OGLE collaboration for the use of facilities of the
Warsaw telescope, for their permanent support and maintenance of the ASAS
instrumentation, and to Carnegie Institution of Washington for providing
the excellent site for the observations.

This work was partly supported by the KBN 2P03D01416 grant.
}

\end{document}

%% file: include.tex
\newcommand{\etal}{{\it et al.\ }}

\newcommand{\beq}{\begin{equation}
  \renewcommand{\int}{\intop\limits}
  \renewcommand{\oint}{\ointop\limits}}
\newcommand{\eeq}{\end{equation}}
\newcommand{\beqarr}{\par\begin{minipage}{11cm} \begin{eqnarray*}}
\newcommand{\eeqarr}{\end{eqnarray*} \end{minipage} \hfill 
   \stepcounter{equation}{\rm (\theequation)}\vspace{3mm}\linebreak}
\newcommand{\bdm}{\begin{displaymath}
  \renewcommand{\int}{\intop\limits}
  \renewcommand{\oint}{\ointop\limits}}
\newcommand{\edm}{\end{displaymath}}

\newcommand{\up}[1]{\ifmmode^{\rm #1}\else$^{\rm #1}$\fi}

\newcommand{\arcd}{\ifmmode^{\circ}\else$^{\circ}$\fi}
\newcommand{\arcm}{\ifmmode{'}\else$'$\fi}
\newcommand{\arcs}{\ifmmode{''}\else$''$\fi}

\newcounter{pagefrom}
\newcounter{pageto}
\newcounter{volume}
\newcounter{year}

\newenvironment{Titlepage}{
\vspace*{2cm}
  \begin{center}
}{
  \end{center}\par\vspace{3mm}
}

\newcommand{\Title}[1]{{\large\bf\boldmath #1 \\[3mm] {\footnotesize by} 
\\[3mm]}}

\newcommand{\Author}[2]{{\large\spaceskip 2pt plus 1pt minus 1pt #1}\\[3mm]
   {\small #2}\\[6mm]}

\newcommand{\Received}[1]{}

\newcommand{\Abstract}[2]{{\footnotesize\begin{center}ABSTRACT\end{center}
\vspace{1mm}\par#1\par
\noindent
{\bf Key words:~~}{\it #2}}}

\newcommand{\FigCap}[1]{\footnotesize\par\noindent Fig.\  % 
  \refstepcounter{figure}\thefigure. #1\par}

\newcommand{\TabCap}[2]{\begin{center}\parbox[t]{#1}{\begin{center}
  \small {\spaceskip 2pt plus 1pt minus 1pt T a b l e}
  \refstepcounter{table}\thetable \\[2mm]
  \footnotesize #2 \end{center}}\end{center}}

\newcommand{\TableFont}{\footnotesize}

\newcommand{\MakeTable}[4]{\begin{table}[htb]\TabCap{#2}{#3}
  \begin{center} \TableFont \begin{tabular}{#1} #4 
  \end{tabular}\end{center}\end{table}}

\newcommand{\MakeTableSep}[4]{\begin{table}[p]\TabCap{#2}{#3}
  \begin{center} \TableFont \begin{tabular}{#1} #4 
  \end{tabular}\end{center}\end{table}}

\newcommand{\MakeTableSepLeft}[4]{\begin{table}[p]\TabCap{#2}{#3}
  \TableFont \hspace*{-2cm}\begin{tabular}{#1} #4 
  \end{tabular}\end{table}}

\newenvironment{references}%
{
\footnotesize \frenchspacing

\renewcommand{\AA}{Astron.\ Astrophys.}

\newcommand{\AJ}{Astron.\ J.}

\newcommand{\Acta}{Acta Astron.}
\newcommand{\MNRAS}{MNRAS}
\renewcommand{\and}{{\rm and }}
\section{{\rm REFERENCES}}
\sloppy \hyphenpenalty10000
\begin{list}{}{\leftmargin1cm\listparindent-1cm
\itemindent\listparindent\parsep0pt\itemsep0pt}}%
{\end{list}\vspace{2mm}}

\def\TYLDA{~}
\newlength{\DW}
\newcommand{\refitem}[5]{\item[]{#1} #2%
\def\REFARG{#3}\ifx\REFARG\TYLDA\else, {\it#3}\fi
\def\REFARG{#4}\ifx\REFARG\TYLDA\else, {\bf#4}\fi
\def\REFARG{#5}\ifx\REFARG\TYLDA\else, {#5}\fi.}

\newcommand{\Section}[1]{\section{\normalsize\bf#1}}

\newcommand{\Acknow}[1]{\par\vspace{5mm}{\bf Acknowledgements.} #1}

%% file: table1.tex
\MakeTableSep{|r|r|r|r|r|r|r|r|r|r|}{10cm}{\label{table1}
Selected Fields observed during the first 2 years of ASAS operation}{
\hline
\multicolumn{1}{|c|}{ID} & \multicolumn{1}{c|}{$\alpha$} & \multicolumn{1}{c|}{$\delta$} & \multicolumn{1}{c|}{$l$} & \multicolumn{1}{c|}{$b$} & \multicolumn{1}{c|}{$N$} & \multicolumn{1}{c|}{$N$} & \multicolumn{1}{c|}{$N$} & \multicolumn{1}{c|}{$f_{per}$} & \multicolumn{1}{c|}{$f_{misc}$}\\
\multicolumn{1}{|c|}{~} & \multicolumn{1}{c|}{2000} & \multicolumn{1}{c|}{2000} & \multicolumn{1}{c|}{~} & \multicolumn{1}{c|}{~} & \multicolumn{1}{c|}{star} & \multicolumn{1}{c|}{per} & \multicolumn{1}{c|}{misc} & \multicolumn{1}{c|}{\%} & \multicolumn{1}{c|}{\%}\\
\hline
Vel & 10:44 & -52:25 & 284.0544 & 5.7212 & 6688 & 14 & 142 & .21 & 2.12\\
Dor & 5:07 & -80:30 & 292.9800 & -30.9557 & 2955 & 4 & 14 & .14 & .47\\
LMC~1 & 5:10 & -68:10 & 278.8949 & -34.4050 & 2680 & 19 & 144 & .71 & 5.37\\
LMC~2 & 5:10 & -70 & 281.0500 & -33.9766 & 2932 & 16 & 153 & .55 & 5.22\\
LMC~3 & 5:40 & -68:10 & 278.3590 & -31.6538 & 2708 & 11 & 179 & .40 & 6.61\\
LMC~4 & 5:40 & -70 & 280.4979 & -31.4555 & 2972 & 6 & 163 & .20 & 5.48\\
Cen~1 & 11:35 & -60 & 293.5079 & 1.4657 & 7444 & 32 & 425 & .43 & 5.71\\
Cen~2 & 11:35 & -61:50 & 294.0422 & -0.2880 & 6716 & 65 & 289 & .96 & 4.30\\
Cen~3 & 11:35 & -63:40 & 294.5766 & -2.0418 & 5640 & 30 & 167 & .53 & 2.96\\
Oct~1 & 12:00 & -85:00 & 301.7311 & -22.2486 & 2004 & 4 & 16 & .20 & .80\\
Vir~1 & 12:30 & 3:00 & 290.0177 & 65.3328 & 572 & 1 & 0 & .17 & 0\\
Cru & 12:50 & -63:00 & 302.7688 & -0.1287 & 5261 & 35 & 214 & .66 & 4.07\\
Vir~2 & 13:25 & -8:50 & 316.8372 & 53.1299 & 715 & 0 & 4 & 0 & .56\\
Cen~4 & 13:50 & -30:00 & 317.7561 & 31.1804 & 1566 & 6 & 18 & .36 & 1.15\\
Cen~5 & 13:50 & -31:50 & 317.2028 & 29.4104 & 1726 & 2 & 11 & .11 & .64\\
Lib & 15:05 & -15:00 & 344.5631 & 36.8518 & 1267 & 4 & 7 & .31 & .55\\
Sgr~1 & 17:00 & -22:30 & 359.5874 & 12.1070 & 3889 & 2 & 177 & .05 & 4.55\\
Sco & 17:55 & -35:00 & 355.7998 & -4.8083 & 8238 & 12 & 758 & .15 & 9.20\\
Sgr~2 & 18:00 & -23:30 & 6.3198 & -0.0102 & 5039 & 18 & 166 & .36 & 3.29\\
Sgr~3 & 19:00 & -22:30 & 13.5471 & -11.8418 & 4678 & 6 & 184 & .19 & 3.93\\
CrA & 19:00 & -40:00 & 356.7925 & -18.5763 & 3263 & 7 & 66 & .21 & 2.02\\
Sgr~4 & 20:00 & -20:30 & 21.2047 & -23.9146 & 2074 & 4 & 16 & .19 & .77\\
Cap~1 & 21:00 & -17:00 & 30.9357 & -35.8809 & 1548 & 2 & 16 & .13 & 1.03\\
Cap~2 & 22:00 & -12:00 & 44.9717 & -47.1328 & 1218 & 1 & 4 & .08 & .33\\
Aqr & 23:00 & -6:30 & 65.7819 & -56.4493 & 1073 & 1 & 3 & .09 & .28\\
Psc & 0:00 & 0:00 & 96.3375 & -60.1886 & 807 & 0 & 1 & 0 & .12\\
S-092 & 0:55 & 1:00 & 124.8203 & -61.8587 & 1075 & 1 & 1 & .09 & .09\\
S-093 & 1:55 & 0:45 & 154.2385 & -58.2047 & 1281 & 1 & 3 & .08 & .23\\
S-094 & 2:56 & 0:30 & 175.3464 & -49.2647 & 1337 & 1 & 2 & .07 & .15\\
S-095 & 3:53 & 0:00 & 188.7133 & -38.6830 & 1466 & 2 & 4 & .14 & .27\\
S-096 & 4:52 & 0:00 & 198.2627 & -26.3019 & 2378 & 8 & 10 & .34 & .42\\
S-097 & 5:57 & 0:00 & 206.6324 & -12.0880 & 3258 & 6 & 38 & .18 & 1.17\\
S-098 & 6:51 & -0:20 & 213.1787 & -0.2493 & 6534 & 31 & 63 & .47 & .96\\
S-099 & 7:56 & -0:20 & 220.7528 & 14.1675 & 3690 & 8 & 20 & .22 & .54\\
S-100 & 8:53 & -0:40 & 228.5439 & 26.4171 & 2065 & 2 & 2 & .10 & .10\\
S-101 & 9:57 & -0:30 & 239.0337 & 39.8217 & 1450 & 2 & 1 & .14 & .07\\
S-102 & 10:57 & -0:20 & 253.1906 & 51.1380 & 1102 & 0 & 4 & 0 & .36\\
S-103 & 11:56 & -0:30 & 274.9302 & 59.3239 & 1052 & 2 & 2 & .19 & .19\\
S-104 & 12:43 & -0:30 & 298.3910 & 62.2973 & 1009 & 1 & 2 & .10 & .20\\
S-105 & 13:37 & -0:20 & 326.5261 & 60.4343 & 1092 & 0 & 1 & 0 & .09\\
S-106 & 14:40 & -0:10 & 351.0974 & 52.2473 & 1268 & 1 & 2 & .08 & .16\\
S-107 & 15:40 & -0:20 & 5.8356 & 41.0914 & 1612 & 2 & 7 & .12 & .43\\
S-108 & 16:37 & 0:20 & 16.3987 & 29.6880 & 2269 & 4 & 7 & .18 & .31\\
S-109 & 17:45 & 0:00 & 25.1863 & 14.7379 & 4359 & 5 & 74 & .11 & 1.70\\
S-110 & 18:41 & 0:10 & 31.8899 & 2.3982 & 3732 & 7 & 124 & .19 & 3.32\\
S-111 & 19:38 & 0:10 & 38.4485 & -10.2678 & 5718 & 8 & 161 & .14 & 2.81\\
S-112 & 20:42 & 0:10 & 46.5206 & -24.3028 & 3125 & 6 & 23 & .19 & .74\\
S-113 & 21:41 & -0:20 & 55.1875 & -37.0604 & 2151 & 2 & 11 & .00 & .51\\
S-114 & 22:40 & 1:00 & 69.2319 & -47.6908 & 1551 & 2 & 5 & .13 & .32\\
S-115 & 23:40 & 1:00 & 88.6300 & -57.0362 & 1354 & 1 & 4 & .07 & .30\\
\hline
}

%% file: table2.tex
\MakeTable{|l|r|r|r|r|r|}{10cm}{\label{table2}
Number of periodic and miscellaneous variable stars detected 
in the selected fields
{\em vs} number of previously known variables in 1 mag bins. Last column lists
a total number of observed stars.}{
\hline
Mag & \multicolumn{2}{c|}{Miscellaneous} & \multicolumn{2}{c|}{Periodic}&Total\\
& $N_{ASAS}$ & $N_{known}$ & ${N_{ASAS}}$ & $N_{known}$ & $N_{obs}$\\
\hline
2-6   & 13 & 8 & 0 & 0 &520\\
6-7   & 12 & 3 & 1 & 0 &570\\
7-8   & 98 & 5 & 10 & 5 &810\\
8-9   & 327 & 17 & 51 & 4 &2500\\
9-10  & 666 & 22 & 12 & 13 &7100\\
10-11 & 1231 & 43 & 148 & 12 &19500\\
11-12 & 904 & 21 & 88 & 8 &39000\\
12-13 & 245 & 2 & 26 & 2 &46100\\
13-   & 5 & 0 & 0 & 0 &16000\\
\hline
}

%% file: table3.tex
\MakeTableSep{|l|r|r|l||l|r|r|l|}{10cm}{\label{table3}
Subset of the ASAS Catalog of Variable Stars - Miscellaneous Variables. Only stars satisfying the condition $\Delta I > 1.5$ and  $I<12$ were included.}{
\hline
\multicolumn{1}{|c|}{ID} & \multicolumn{1}{c|}{$I$-mag} & \multicolumn{1}{c|}{$\Delta~I$} & \multicolumn{1}{c||}{GCVS} & \multicolumn{1}{|c|}{ID} & \multicolumn{1}{c|}{$I$-mag} & \multicolumn{1}{c|}{$\Delta~I$} & \multicolumn{1}{c|}{GCVS}\\
\hline
$050135-6805.9$ & 8.324 & 2.172 & RX~DOR & $064728-0101.5$ & 9.704 & 2.050 & ~\\
$075743-0041.1$ & 9.119 & 1.765 & ~ & $075939+0030.5$ & 10.147 & 2.230 & AF~CMI\\
$085816-0049.9$ & 7.787 & 2.472 & ~ & $103706-5313.7$ & 10.354 & 2.042 & ~\\
$103817-5136.1$ & 10.137 & 2.365 & ~ & $104317-5333.3$ & 8.637 & 1.987 & ~\\
$104448-5225.4$ & 9.754 & 5.517 & ~ & $104956-5246.1$ & 8.092 & 2.424 & ~\\
$105341-5311.5$ & 7.461 & 1.717 & RU~VEL & $105441-5203.8$ & 9.949 & 1.811 & ~\\
$110055-8405.4$ & 10.410 & 1.823 & ~ & $112407-6146.0$ & 9.421 & 2.853 & ~\\
$112638-5947.8$ & 9.652 & 1.968 & V780~CEN & $112939-6011.0$ & 9.962 & 1.707 & ~\\
$113321-6135.4$ & 10.414 & 2.576 & V781~CEN & $113632-6257.4$ & 8.930 & 1.920 & ~\\
$114107-6129.3$ & 10.396 & 1.710 & ~ & $114130-6004.9$ & 10.768 & 1.811 & ~\\
$114523-5942.4$ & 9.455 & 2.083 & ~ & $114630-6153.4$ & 7.986 & 2.431 & VV~CEN\\
$114754-6344.9$ & 9.904 & 2.125 & ~ & $114857-6404.6$ & 8.499 & 1.742 & ~\\
$130330-6339.5$ & 9.723 & 1.906 & TY~CEN & $135333-2905.6$ & 10.242 & 1.665 & ~\\
$135743-3104.1$ & 5.796 & 2.798 & TW~CEN & $150855-1529.8$ & 8.210 & 2.721 & TT~LIB\\
$164153+0101.1$ & 10.709 & 2.050 & ~ & $170002-2325.3$ & 10.531 & 1.592 & V2097~OPH\\
$170150-2311.6$ & 9.645 & 2.645 & V2098~OPH & $170323-2316.3$ & 9.965 & 2.382 & ~\\
$170353-2144.4$ & 10.439 & 2.717 & V1290~OPH & $170601-2157.6$ & 9.939 & 2.626 & V1353~OPH\\
$170630-2242.0$ & 10.821 & 1.756 & ~ & $174428+0049.1$ & 9.564 & 1.773 & V376~OPH\\
$174651+0056.2$ & 9.910 & 1.971 & ~ & $174803-0041.0$ & 8.235 & 1.803 & ~\\
$174820-3542.1$ & 7.600 & 1.595 & SV~SCO & $174843-3444.3$ & 10.665 & 2.732 & ~\\
$174947-3505.9$ & 10.874 & 1.904 & V397~SCO & $175210-3542.2$ & 10.261 & 1.626 & ~\\
$175225-3503.3$ & 8.624 & 4.255 & V407~SCO & $175226-3411.4$ & 10.935 & 2.542 & ~\\
$175340-3401.0$ & 10.132 & 1.935 & ~ & $175342-3556.3$ & 10.563 & 1.660 & ~\\
$175349-3424.0$ & 7.864 & 2.437 & SY~SCO & $175358-3423.6$ & 10.344 & 1.942 & V420~SCO\\
$175410-3420.5$ & 6.240 & 1.827 & BN~SCO & $175430-3548.9$ & 10.720 & 1.944 & V421~SCO\\
$175527-2406.8$ & 8.910 & 2.805 & ~ & $175651-3509.2$ & 10.714 & 1.734 & ~\\
$175712-3541.7$ & 10.632 & 1.658 & ~ & $175847-3538.4$ & 10.581 & 3.475 & ~\\
$175930-3539.6$ & 10.999 & 1.648 & ~ & $175958-3523.2$ & 10.333 & 2.965 & V537~SGR\\
$180021-3456.7$ & 9.880 & 2.163 & V544~SGR & $180032-3438.6$ & 8.500 & 1.952 & FT~SGR\\
$180033-2416.3$ & 8.495 & 2.367 & ~ & $180043-3520.7$ & 10.366 & 2.189 & V549~SGR\\
$180100-3436.7$ & 10.444 & 2.119 & FU~SGR & $180156-3556.8$ & 10.313 & 1.848 & ~\\
$180157-3511.6$ & 8.981 & 3.498 & V561~SGR & $180220-3454.2$ & 10.793 & 3.642 & FV~SGR\\
$180240-2322.0$ & 10.687 & 1.973 & V1950~SGR & $180306-3436.2$ & 10.454 & 2.082 & ~\\
$180331-2259.7$ & 9.677 & 1.834 & V1951~SGR & $180502-2327.6$ & 10.925 & 2.545 & ~\\
$180657-2314.9$ & 10.605 & 1.727 & ~ & $184517+0056.5$ & 10.902 & 1.887 & ~\\
$185349-3937.9$ & 10.636 & 2.418 & ~ & $185357-2316.4$ & 10.872 & 2.268 & V2062~SGR\\
$185449-2334.7$ & 10.678 & 2.272 & ~ & $185457-2331.7$ & 10.684 & 2.019 & ~\\
$185509-2246.8$ & 10.668 & 1.695 & ~ & $185614-2132.8$ & 10.578 & 1.999 & ~\\
$185654-3925.6$ & 9.950 & 1.621 & AB~CRA & $185802-2234.6$ & 8.264 & 2.725 & BU~SGR\\
$185933-2206.1$ & 10.457 & 2.722 & V2083~SGR & $190049-2134.5$ & 9.525 & 2.463 & V2090~SGR\\
$190303-3942.9$ & 9.341 & 2.836 & ~ & $190333-2317.8$ & 9.527 & 1.598 & ~\\
$190559-2207.7$ & 10.198 & 2.379 & ~ & $190659-2244.3$ & 10.573 & 2.484 & V337~SGR\\
$193200-0028.6$ & 10.925 & 2.555 & ~ & $193526+0042.7$ & 8.226 & 2.649 & V607~AQL\\
$193956-0035.9$ & 9.089 & 3.207 & ~ & $194007-0020.8$ & 8.112 & 1.709 & ~\\
$204637-0054.1$ & 8.324 & 2.153 & ~ & $214805+0022.1$ & 10.164 & 2.422 & TY~AQR\\
\hline
}

%% file: table4.tex
\MakeTableSepLeft{|l|r|r|r|l||l|r|r|r|l|}{10cm}{\label{table4}
Subset of the ASAS Catalog of Variable Stars - Periodic Variables. Only stars satisfying the condition $\Delta I > 0.3$ and  $I<12$ were included.}{
\hline
\multicolumn{1}{|c|}{ID} & \multicolumn{1}{c|}{$I$-mag} & \multicolumn{1}{c|}{$\Delta~I$} & \multicolumn{1}{c|}{$P$} & \multicolumn{1}{c||}{GCVS} & \multicolumn{1}{c|}{ID} & \multicolumn{1}{c|}{$I$-mag} & \multicolumn{1}{c|}{$\Delta~I$} & \multicolumn{1}{c|}{$P$} & \multicolumn{1}{c|}{GCVS}\\
\hline
$045206-7043.9$ & 10.500 & 0.700 & 4.68927 & ~ & $045423-7054.1$ & 11.722 & 0.522 & 34.5602 & ~\\
$045720-8023.0$ & 11.529 & 0.714 & .366925 & ~ & $045810-6957.0$ & 11.801 & 0.600 & 39.4135 & ~\\
$045832-7020.8$ & 11.927 & 0.585 & 35.7411 & ~ & $050047-7029.8$ & 10.683 & 0.355 & .387359 & ~\\
$050648-7002.2$ & 11.845 & 0.628 & 47.4203 & ~ & $050720-7027.2$ & 12.604 & 0.941 & 26.2895 & ~\\
$050818-6846.8$ & 11.492 & 0.326 & 30.4386 & ~ & $050920-7027.4$ & 11.847 & 0.637 & 37.4883 & ~\\
$051023-6846.4$ & 10.394 & 0.351 & 74.4470 & ~ & $051833-6813.6$ & 10.027 & 0.548 & .285465 & RW~DOR\\
$051931-6841.2$ & 10.977 & 0.453 & 118.73 & ~ & $052507-6738.6$ & 11.283 & 0.571 & 48.2742 & ~\\
$052650-8135.2$ & 7.913 & 0.430 & .461666 & TY~MEN & $053120-7057.5$ & 11.539 & 0.464 & 52.3697 & ~\\
$053959-6828.7$ & 10.713 & 0.484 & .362219 & ~ & $055122-6812.8$ & 11.736 & 0.375 & .321769 & ~\\
$064558-0017.5$ & 10.062 & 0.552 & .568018 & ~ & $064918-0003.5$ & 12.371 & 0.694 & .773591 & V514~MON\\
$064919+0019.8$ & 10.725 & 0.505 & 2.59701 & V450~MON & $065032+0000.4$ & 10.939 & 0.344 & 7.09559 & TW~MON\\
$065052-0125.8$ & 9.650 & 0.417 & 8.70508 & TX~MON & $065128-0122.3$ & 9.109 & 0.972 & 16.3744 & SZ~MON\\
$065144-0034.6$ & 11.970 & 0.575 & 1.22343 & ~ & $065640+0011.5$ & 10.347 & 0.361 & 4.02311 & TY~MON\\
$075222-0117.5$ & 11.709 & 0.832 & 5.75735 & ~ & $075623-0043.7$ & 11.285 & 0.426 & 1.15097 & ~\\
$075717-0005.0$ & 11.538 & 0.446 & .660746 & ~ & $075947+0021.0$ & 11.545 & 0.553 & 1.05386 & ~\\
$103617-5202.5$ & 11.965 & 0.409 & .422209 & ~ & $104526-5224.3$ & 10.308 & 0.683 & 2.38364 & FW~VEL\\
$104528-5321.9$ & 10.830 & 0.523 & 95.3013 & ~ & $104755-5214.9$ & 6.553 & 0.554 & 1.76740 & ~\\
$112404-6403.2$ & 11.039 & 0.421 & 1.70404 & ~ & $112612-6210.2$ & 10.533 & 0.391 & 1.31624 & ~\\
$112645-6251.8$ & 11.188 & 0.479 & 1.59504 & ~ & $112653-6211.9$ & 11.999 & 0.675 & .587651 & V343~CEN\\
$112803-6124.7$ & 8.609 & 0.483 & 3.48927 & MN~CEN & $112927-6201.9$ & 10.577 & 0.321 & 3.2246 & ~\\
$112939-5953.7$ & 11.487 & 0.367 & 1.91642 & ~ & $113318-6314.6$ & 10.679 & 0.391 & 4.97437 & LV~CEN\\
$113617-6128.1$ & 8.619 & 0.892 & 3.69340 & BF~CEN & $113708-6148.0$ & 10.140 & 0.381 & 1.22635 & ~\\
$114346-6144.6$ & 9.729 & 0.341 & 2.99289 & MP~CEN & $114353-6024.7$ & 10.090 & 0.343 & 3.22799 & ~\\
$114510-6058.2$ & 10.154 & 0.464 & 3.91355 & MR~CEN & $114556-5922.6$ & 11.417 & 0.695 & .453195 & ~\\
$114726-6132.9$ & 11.792 & 0.444 & 78.3978 & ~ & $114757-6033.9$ & 8.810 & 0.926 & 1.65749 & ~\\
$115934-8546.0$ & 11.222 & 0.328 & .611267 & ~ & $123808-6353.8$ & 11.492 & 0.563 & 1.12587 & VZ~CRU\\
$124203-6226.2$ & 10.896 & 0.525 & 1.90701 & ~ & $124422-6300.7$ & 11.375 & 0.550 & 12.6528 & ~\\
$124715-6309.8$ & 11.125 & 0.919 & 122.49 & ~ & $125159-6353.2$ & 11.888 & 0.669 & 29.6164 & ~\\
$125420-6356.1$ & 10.610 & 0.341 & 24.0326 & ~ & $125427-6356.1$ & 9.895 & 0.656 & 24.0462 & RY~CRU\\
$125816-6258.1$ & 11.093 & 0.515 & 2.52430 & ~ & $125900-6405.0$ & 11.942 & 0.358 & 8.08679 & ~\\
$135336-2934.7$ & 11.952 & 0.556 & .636646 & FY~HYA & $164121+0030.4$ & 7.842 & 0.422 & .453391 & V502~OPH\\
$164254+0107.4$ & 10.678 & 1.095 & 151.70 & ~ & $174108+0029.8$ & 9.834 & 2.259 & 172.45 & ~\\
$175147-3500.7$ & 11.673 & 0.504 & 116.373 & ~ & $175322-3550.9$ & 10.832 & 1.565 & 182.31 & ~\\
$180059-2301.9$ & 9.224 & 0.836 & 4.67002 & WY~SGR & $180209-3429.6$ & 10.781 & 0.341 & 27.4366 & V564~SGR\\
$180253-2409.6$ & 9.797 & 0.445 & 2.10910 & ~ & $180305-2251.8$ & 9.507 & 0.764 & 3.93154 & V792~SGR\\
$180326-2236.9$ & 8.098 & 0.442 & 1.39134 & ~ & $180328-2332.1$ & 11.300 & 0.546 & 1.70632 & ~\\
$180428-2260.0$ & 11.561 & 0.556 & 6.69200 & ~ & $180445-2243.8$ & 10.595 & 0.345 & 15.4463 & ~\\
$180449-2243.9$ & 8.830 & 0.692 & 15.3882 & AV~SGR & $184010-0047.7$ & 10.814 & 0.768 & .435804 & ~\\
$184139-0044.7$ & 11.031 & 0.325 & .287523 & ~ & $185448-2326.2$ & 10.415 & 0.445 & 3.52499 & ~\\
$185627-4043.9$ & 11.756 & 0.384 & .360170 & ~ & $190403-2228.6$ & 11.808 & 0.708 & 1.35612 & V1071~SGR\\
$190514-2316.5$ & 11.942 & 0.402 & 1.28938 & ~ & $193459-0032.4$ & 9.997 & 0.869 & 201.00 & ~\\
$193640+0053.7$ & 11.016 & 0.335 & 2.00165 & V1269~AQL & $200339-1956.0$ & 11.054 & 0.321 & .912107 & ~\\
$203901+0056.0$ & 10.372 & 2.547 & 304.38 & ~ & $204045+0056.4$ & 8.270 & 0.369 & 2.36813 & ~\\
$204859+0027.4$ & 11.133 & 0.408 & .513487 & ~ & $210554-1647.8$ & 11.777 & 0.333 & .303387 & ~\\
$214610-0106.8$ & 11.768 & 0.589 & .285142 & ~ & $224553+0102.9$ & 10.280 & 0.545 & .721005 & DD~AQR\\
\hline
}

%% file: asas3.bbl
\begin{references}
\refitem{Landolt, A.U.}{1992}{\AJ}{104}{340}
\refitem{Kholopov, P.N., \etal}{1985}{~}{~}{General Catalog of Variable 
Stars, The Fourth Edition, Nauka, Moscow} 
\refitem{Paczy{\'n}ski, B.}{1997}{~}{~}{"The Future of Massive Variability 
Searches", in {\it Proceedings of 12th IAP Colloquium}: "Variable Stars and the 
Astrophysical Returns of Microlensing Searches", Paris (Ed. R. Ferlet), 
p.~357} 
\refitem{Perryman, M.A.C. \etal}{1997}{\AA}{323}{L49}
\refitem{Pojma{\'n}ski, G.}{1997}{\Acta}{47}{467}
\refitem{Pojma{\'n}ski, G.}{1998}{\Acta}{48}{35}
\refitem{Schwarzenberg-Czerny, A.}{1989}{\MNRAS}{241}{153}
\refitem{Udalski, A. Kubiak, M., and Szyma{\'n}ski, M.}{1997}{\Acta}{47}{319}
\end{references}
